\documentclass[12pt]{article}
\usepackage{latexsym}
\begin{document}

\title{A Lorentz invariant formulation of the Yang-Mills theory with gauge invariant ghost field Lagrangian.}
\author{A.A.Slavnov \\ Steklov Mathematical Institute\\ and Moscow State University \\ Moscow}

\maketitle

\begin{abstract}
A new formulation of the Yang-Mills theory which allows a manifestly covariant gauge
fixing accompanied by a gauge invariant ghost field interaction is proposed. The gauge
condition selects a unique representative in the class of gauge equivalent
configurations.
\end{abstract}

\section{Introduction.}

Quantization of the Yang-Mills theory involves fixing a gauge. A
Lorentz invariant gauge fixing in non-Abelian theories requires
introduction of additional anticommuting scalar fields,
Faddeev-Popov ghosts \cite{FP}, with non gauge invariant
interaction. The ghost field interaction produces ultraviolet
divergencies which has to be removed by renormalization of the ghost
fields and the ghost-gluon interaction vertex. Moreover such a gauge
fixing does not choose a unique representative in a class of gauge
equivalent configurations. For large fields Gribov copies appear,
which makes questionable using this procedure as a starting point
for nonperturbative calculations. Contrary to the Quantum
Electrodynamics (QED), where the quantization procedure does not
break the conservation of the current interacting with the gauge
field, in non-Abelian theories this conservation is broken both by
the gauge fixing and by the presence of non-gauge invariant ghost
field interaction. The Ward identities
(\cite{Wa},\cite{Fr},\cite{Ta}) which express in the quantum case
the conservation of electromagnetic current are replaced in the
non-Abelian theories by much more complicated relations,
Slavnov-Taylor (ST) identities (\cite{Sl1},\cite{T}), including the
Green functions of composite operators. These relations may be
interpreted as a consequence of the conservation of a more general
current involving also the ghost fields
(\cite{BRS1},\cite{BRS2},\cite{Ty}).

Dynamical ghost fields are absent in the linear gauges, like $nA=0$, which makes
possible to obtain in this case relations between the Green functions, similar to the
Ward identities in QED. However linear gauges break explicitly the Lorentz invariance,
and their using is very cumbersome. According to the common wisdom in non-Abelian
theories one has to sacrifice either explicit Lorentz invariance or gauge invariance
of the ghost field action.

In this paper I propose a procedure which preserves in the quantum Yang-Mills theory
simultaneously the manifest Lorentz invariance and the gauge invariance of the ghost
field Lagrangian. It allows to obtain the relations between the Green functions
following essentially the same procedure as in the Abelian case. The gauge fixing used
in the present paper is free of Gribov ambiguity and in perturbation theory leads to
the same results as the standard quantization procedure.

The paper is organized as follows. In the next section the model is described and its
equivalence to the standard Yang-Mills theory is demonstrated. In the third section a
new Lorentz invariant gauge condition free of Gribov ambiguity is introduced and the
diagram technique is analyzed. The relations between the Green functions are derived.
In Conclusion I discuss the results obtained in this paper and their possible
applications.

\section{The model.}

In this section I consider the $SU(2)$ gauge model. Generalization to other compact
groups does not make serious problems.

We start with the usual path integral representation for the $S$-matrix in the Coulomb
gauge
\begin{equation}
S=\int\exp\{i\int[L_{YM}+\lambda^a\partial_iA_i^a]dx\}d\mu \label{1}
\end{equation}
where
\begin{equation}
L_{YM}=- \frac{1}{4}F^a_{\mu\nu}F^a_{\mu\nu} \label{2}
\end{equation}
\begin{equation}
F^a_{\mu\nu}=\partial_{\mu}A_{\nu}^a-\partial_{\nu}A_{\mu}^a+g\varepsilon^{abc}A_{\mu}^bA_{\nu}^c,
\quad a,b,c,=1,2,3\label{3}
\end{equation}
The measure $d\mu$ includes differentials of all the fields as well as the
Faddeev-Popov determinant $\det{M}$. This determinant is conveniently presented as the
integral over anticommuting ghost fields
\begin{equation}
\det{M}=\int\exp\{i\int \bar{c}\partial_iD_ic dx\}d \bar{c}dc
 \label{3a}
\end{equation}
where $D_i$ is a covariant derivative.

The effective action in the integral (\ref{1}) is not gauge invariant. Contrary to the
Abelian case the gauge invariance is broken not only by the gauge fixing term but also
by the Faddeev-Popov ghost Lagrangian. To avoid this complication I propose the
following construction.

Let us consider the path integral
\begin{eqnarray}
S= \int \exp\lbrace i \int [L_{YM}+(D_{\mu}\varphi)^*(D_{\mu}\varphi)-(D_{\mu}\chi)^*(D_{\mu}\chi)+\nonumber\\
 (D_{\mu}b)^*(D_{\mu}e)+(D_{\mu}e)^*(D_{\mu}(b)]dx \delta(\partial_iA_i)d\mu'
\label{4}
\end{eqnarray}
The measure $d\mu'$ differs of $d\mu$ by the product of differentials of the scalar
fields $(\varphi,\varphi^*,\chi,\chi^*,b,b^*,e,e^*)$.

We assume that the scalar fields comprise complex $SU(2)$ doublets, the fields
$\varphi,\chi$ are commuting and $d,e$ are anticommuting. The integration goes over
the scalar fields with radiation (Feynman) boundary conditions, which corresponds to
considering the matrix elements between states which do not include excitations
corresponding to the scalar ghost fields. The gauge fields in the integral (\ref{4})
satisfy the boundary conditions
\begin{equation}
A_i^{tr}\rightarrow A_i^{tr}(\emph{in,out}), \quad t\rightarrow\mp\infty \label{5}
\end{equation}
where $A_i^{tr}$ are the three dimensionally transversal components of $A_i$, all
other fields having vacuum boundary conditions. Obviously due to the presence of
$\delta(\partial_iA_i)$, $\partial_i A_i=0$ at any time.

Performing explicitly the integration over the scalar fields in the eq.(\ref{4}), we
get the factor $(|D|^2)^{-2}$ from the integration over commuting fields $\varphi$ and
$\chi$, and the factor $(|D|^2)^2$ from the integration over the anticommuting fields
$b$ and $e$. Hence the integral (\ref{4}) coincides with the integral (\ref{1}), which
justifies the using in the l.h.s. of eq.(\ref{4}) the same symbol $S$.

Let us make in the integral (\ref{4}) the shift of the integration variables
\begin{equation}
\varphi \rightarrow \varphi+g^{-1} \hat{a}, \quad \chi \rightarrow \chi-g^{-1}
\hat{a}, \quad \hat{a}=(0,\frac{a}{ \sqrt{2}})
 \label{5a}
\end{equation}
$a$ is a constant parameter. Instead of the eq.(\ref{4}) now we have
\begin{eqnarray}
\tilde{S}= \int \exp \{ \int[L_{YM}+(D_{\mu}\varphi)^*(D_{\mu}\varphi)
+g^{-1}(D_{\mu}\varphi)^*(D_{\mu} \hat{a})+\nonumber\\
g^{-1}(D_{\mu} \hat{a})^*(D_{\mu}\varphi)+g^{-1}(D_{\mu}\chi)^*(D_{\mu} \hat{a})
+g^{-1}(D_{\mu} \hat{a}^*)(D_{\mu}\chi)+\nonumber\\
-(D_{\mu}\chi)^*(D_{\mu}\chi)+(D_{\mu}b)^*(D_{\mu}e)
+(D_{\mu}e)^*(D_{\mu}b)]dx\}\delta(\partial_iA_i)d\mu' \label{6}
\end{eqnarray}
Although at first sight the transformation (\ref{5a}) may influence the asymptotic
behavior of the integration variables and therefore change the value of the integral,
in our case it does not happen. Indeed making in the integral (\ref{6}) the
transformation
\begin{eqnarray}
\varphi(x)= \varphi'(x)-g^{-1} \int D^{-2}(x,y)(D^2 \hat{a})(y)dy\nonumber\\
\chi(x)= \chi'(x)+g^{-1} \int D^{-2}(x,y)(D^2 \hat{a})(y)dy \label{7}
\end{eqnarray}
which is a legitimate change of variables as $D^2 \hat{a}$ is decreasing fast at $|t|
\rightarrow \infty$, we are coming back to the eq.(\ref{4}). (Note that it is
impossible to integrate in the eq.(\ref{7}) by parts, as $\hat{a}$ is a constant
spinor.)

The action in the exponent (\ref{6}) is invariant with respect to "shifted" gauge
transformations
\begin{eqnarray}
\delta A_{\mu}^a=\partial_{\mu}\eta^a-g \varepsilon^{abc}A_{\mu}^b \eta^c \nonumber\\
\delta \varphi^0= \frac{g}{2} \varphi^a \eta^a \nonumber\\
\delta\varphi^a=- \frac{a \eta^a}{2}-\frac{g}{2}\varepsilon^{abc}\varphi^b\eta^c-\frac{g}{2}\varphi^0 \eta^a \nonumber\\
 \delta \chi^a= \frac{a \eta^a}{2}- \frac{g}{2} \varepsilon^{abc} \chi^b \eta^c-
\frac{g}{2} \chi^0 \eta^a \nonumber\\
\delta \chi^0=\frac{g}{2} \chi^a \eta^a \nonumber\\
\delta b^a=-\frac{g}{2}\varepsilon^{adc}b^d \eta^c-\frac{g}{2}b^0 \eta^a \nonumber\\
\delta b^0= \frac{g}{2} b^a \eta^a \nonumber\\
\delta e^a=-\frac{g}{2}\varepsilon^{adc}e^d \eta^c \nonumber\\
\delta e^0= \frac{g}{2}e^a \eta^a, \label{8}
\end{eqnarray}
where we introduced the representations of the scalar fields in terms of Hermitian
components, e.g.
\begin{equation}
\varphi=(\frac{i \varphi_1+\varphi_2}{\sqrt{2}},
\frac{\varphi_0-i\varphi_3}{\sqrt{2}})
\label{9}
\end{equation}

This action is also invariant with respect to the supersymmetry transformations
\begin{eqnarray}
 \delta \varphi(x)= \epsilon b(x)\nonumber\\
\delta \chi(x)=- \epsilon b(x) \nonumber\\
\delta e(x)=\epsilon (\varphi(x)+\chi(x)) \nonumber\\
\delta b=0 \label{10}
\end{eqnarray}
where $\epsilon$ is an anticommuting constant parameter. This invariance is closely
related to the fact that the integral (\ref{4}) in the sector which does not contain
excitations corresponding to the scalar fields coincides with the Yang-Mills
scattering matrix (see \cite{Sl2}, \cite{Sl3}, \cite{Sl4}).

Our proof of equivalence of representations (\ref{4}) and (\ref{6}) did not take into
account a necessity of renormalization. One can see that to remove ultraviolet
infinities generated in the perturbative expansion of the integral (\ref{6}) mass
renormalization of the type
\begin{equation}
 \delta m_g(e^*d+d^*e), \quad \delta m_{\varphi} \varphi^* \varphi, \quad \delta
m_{\chi} \chi^* \chi \label{11}
\end{equation}
may be needed, as well as new four point vertices
\begin{equation}
\gamma(e^*b+b^*e)^2, \quad \mu(\varphi^* \varphi)^2, \quad \varrho(\chi^* \chi)^2
\label{12}
\end{equation}
A possible counterterm structures not present in the Lagrangian (\ref{6}) and
compatible with the symmetries (\ref{8}), (\ref{10}) are
\begin{eqnarray}
A[b^*e+e^*b+ \varphi^* \varphi- \chi^* \chi+a(\varphi_0+\chi_0)]\nonumber\\
B[b^*e+e^*b+\varphi^* \varphi-\chi^* \chi+a(\varphi_0+\chi_0)]^2 \label{13}
\end{eqnarray}
where $A \sim g^2$, $B\sim g^4$. Note that the terms linear in the fields $\varphi,
\chi$ are present in the eq.(\ref{13}), corresponding to the necessity of the tadpole
renormalization. In general any counterterms compatible with the symmetries of the
theory may arise. In the presence of such terms in the exponent of the integral
(\ref{6}) one cannot any more to prove the equivalence of this expression to the
original representation (\ref{4}) for the Yang-Mills scattering matrix by simple shift
of integration variables. Nevertheless the invariance of the action in the exponent
(\ref{6}) with respect to the transformations (\ref{8}, \ref{10}) provides the
unitarity of the $S$-matrix (\ref{6}) in the physical subspace which includes only
transversal spin one excitations. The proof goes in analogy with the construction
given in the papers (\cite{Sl2}, \cite{Sl3}, \cite{Sl4}).

The invariance of the action with respect to the supersymmetry transformations
(\ref{10}) leads to existence of the conserved charge $Q$ and one can separate the
physical subspace by requiring its annihilation by the charge $Q$. For asymptotic
states we shall have
\begin{equation}
Q^0|\psi>^{as}_{ph}=0
 \label{14}
\end{equation}
where $Q^0$ is the asymptotic conserved charge. The asymptotic charges has a form
\begin{equation}
Q^0 \sim \int[\partial_0 b^{\alpha}(\varphi+
\chi)^{\alpha}-\partial_0(\varphi+ \chi)^{\alpha}b^{\alpha}]d^3x
\label{15}
\end{equation}
I recall that we are working in perturbation theory and assume that the interaction is
asymptotically turned off. That means all the terms $\sim g$ do not contribute to the
asymptotic charge. Being written in terms of creation and annihilation operators the
asymptotic charge looks as follows
\begin{equation}
Q^0 \sim
\int[a_b^{\alpha+}(a_{\chi}^{\alpha-}+a_{\varphi}^{\alpha-})+(a_{\chi}^{\alpha+}+a_{\varphi}^{\alpha+})a_b^{\alpha-}]d^3k
\label{16}
\end{equation}
where the operators $a^{\pm}$ satisfy the following (anti)commutation relations
\begin{eqnarray}
a_b^{\alpha-}(k)a_e^{\beta+}(k')+a_e^{\beta+}(k')a_b^{\alpha-}(k)= \delta^{\alpha\beta} \delta(k-k')\nonumber\\
a_e^{\alpha-}(k)a_b^{\beta+}(k')+a_b^{\beta+}(k')a_e^{\alpha-}(k)=
\delta^{\alpha\beta} \delta(k-k') \label{17}
\end{eqnarray}
\begin{eqnarray}
a_{\varphi}^{\alpha-}(k)a_{\varphi}^{\beta+}(k')-a_{\varphi}^{\beta+}(k')a_{\varphi}^{\alpha-}(k)=
\delta^{\alpha\beta} \delta(k-k')\nonumber\\
a_{\chi}^{\alpha-}(k)a_{\chi}^{\beta+}(k')-a_{\chi}^{\beta+}(k')a_{\chi}^{\alpha-}(k)=-
\delta^{\alpha\beta} \delta(k-k') \label{17a}
\end{eqnarray}

The operator $Q^0$ is obviously nilpotent as the operators $a_b^+,a_b^-$ are
anticommuting and the operators $(a_{\chi}^-+a_{\varphi}^-),
(a_{\chi}^++a_{\varphi}^+)$ are mutually commuting.

Nonnegativity of the subspace annihilated by the operator $Q^0$ may be proven in the
usual way (see \cite{He}). Introducing the number operator for unphysical scalar modes
\begin{equation}
\hat{N}=\int
\{a_{\varphi}^+(\textbf{k})a_{\varphi}^-(\textbf{k})-a_{\chi}^+(\textbf{k})a_{\chi}^-
((\textbf{k})+a_b^+(\textbf{k})a_e^-(\textbf{k})+a_e^+(\textbf{k})a_b^-(\textbf{k})\}d^3k
\label{18}
\end{equation}
we see that this operator may be presented as the anticommutator
\begin{equation}
\hat{N}=[Q^0,K^0]_+
 \label{19}
\end{equation}
where
\begin{equation}
K^0=\int\{a_e^+(\textbf{k})(a_{\chi}^-(\textbf{k})-a_{\varphi}^-(\textbf{k}))+(a_{\chi}^+(\textbf{k})
-a_{\varphi}^+(\textbf{k}))a_e^-(\textbf{k})\}d^3k \label{20}
\end{equation}
Applying the number operator (\ref{18}) to an arbitrary vector we get
\begin{equation}
\hat{N}|\psi>=N|\psi>
 \label{21}
\end{equation}
if $N \neq 0$ it follows that
\begin{equation}
N|\psi>=Q^0K^0|\psi>+K^0Q^0|\psi>
 \label{22}
\end{equation}
and any vector annihilated by $Q^0$ has a form
\begin{equation}
|\tilde{\psi}>=|\psi>_{A,c}+Q^0|\omega>
 \label{23}
\end{equation}
 where $|\psi>_{A,c}$ does not contain the excitations corresponding to the ghost fields $\varphi,
 \chi,b,e$. Recollecting that this vector contains only three dimensionally transversal excitations
 of the Yang-Mills field we conclude that
\begin{equation}
|\psi>^{as}_{ph}=|\psi>_{tr}+|N>
 \label{24}
\end{equation}
Here the vector $|\psi>_{tr}$ depends only on the three dimensionally transversal
Yang-Mills field excitations and $|N>$ is a zero norm vector orthogonal to
$|\psi>_{tr}$. Factorizing this subspace with respect to the vectors $|N>$ we see that
the $S$-matrix (\ref{6}) with the counterterms respecting the gauge invariance
(\ref{8}) and supersymmetry (\ref{10}) is unitary in the subspace which contains only
three dimensionally transversal excitations of the Yang-Mills field.

\section{An unambiguous Lorentz covariant gauge with gauge invariant ghost interaction.}

Up to now we considered the Yang-Mills theory in the Coulomb gauge and our
reformulation did not give any advantages in comparison with the standard one. In
particular non gauge invariant interaction of the Faddeev-Popov ghosts was present.
However we may pass in the integral (\ref{6}) to some other gauge and get rid off the
 non gauge invariant ghost field Lagrangian. The new gauge condition avoids the problem
 of existence of Gribov copies for large fields.

We consider the gauge
\begin{equation}
\varphi^a- \chi^a=0
 \label{25}
\end{equation}
Obviously this condition selects a unique representative in the
class of gauge equivalent configurations.

To pass to this gauge we shall use the standard Faddeev-Popov trick, multiplying the
integral (\ref{6}) by "one"
\begin{equation}
\Delta \prod_a \int \delta(\varphi^{\Omega}-\chi^{\Omega})^a d\Omega
 \label{26}
\end{equation}
At the surface $\varphi^a- \chi^a=0$ the gauge invariant functional $\Delta$ is equal
to
\begin{equation}
 \Delta^{-1}=\prod_x(a+ \frac{g}{2}(\varphi^0-\chi^0))^{-3}
\label{27}
\end{equation}
Hence in the gauge (\ref{25}) the $S$-matrix generating functional
may be written as follows
\begin{eqnarray}
S=\int
\exp\{i\int[L_{YM}+(D_{\mu}\varphi)^*(D_{\mu}\varphi)-(D_{\mu}\chi)^*(D_{\mu}\chi)
\nonumber\\
+g^{-1}[(D_{\mu}\varphi)^*+(D_{\mu}\chi)^*](D_{\mu} \hat{a})+g^{-1}(D_{\mu} \hat{a})^*
(D_{\mu}\varphi+D_{\mu}\chi)+(D_{\mu}b)^*(D_{\mu}e)+\nonumber\\(D_{\mu}e)^*(D_{\mu}b)
+\lambda^a(\varphi^a-\chi^a)]dx\}\Delta d \tilde{\mu} \label{28}
\end{eqnarray}
The measure $d \tilde{\mu}$ is the product of all the fields differentials and does
not include any dynamical ghost determinants. All the terms in the exponent except for
the gauge fixing term $ \int \lambda^a(\varphi^a-\chi^a)dx$ are invariant with respect
to the gauge transformations (\ref{8}).

The integral (\ref{28}) includes a local measure, which may be formally presented as
an addition to the action having a form
\begin{equation}
\delta A= \int \delta^4(0)\ln(1+\frac{g(\varphi^0-\chi^0)^3}{2a})d^4x \label{29}
\end{equation}
This term compensates some ultraviolet divergencies present in the diagrams generated
by the expansion of the integral (\ref{28}). We shall not analyze this cancelation in
details and assume that this integral is calculated by using a regularization similar
to the dimensional one, that is we omit all counterterms proportional to $\delta(0)$
or $D_c(0)$.

The free action determining the propagators for the perturbative expansion of the
integral (\ref{28}) looks as follows
\begin{eqnarray}
A_0=\int [- \frac{1}{4}(\partial_{\mu}A_{\nu}-\partial_{\nu}A_{\mu})^2+
\frac{1}{2}\partial_{\mu}\varphi^0 \partial_{\mu}\varphi^0- \nonumber\\
\frac{1}{2} \partial_{\mu}\chi^0 \partial_{\mu}\chi^0+a
\partial_{\mu}\varphi^aA_{\mu}^a+ \frac{1}{2}\partial_{\mu}b^{\alpha}\partial_{\mu}e^{\alpha}]dx
\label{30}
\end{eqnarray}
where we used the gauge condition $\varphi^a=\chi^a$.

One sees that the propagators $ \varphi^0,\varphi^0;  \chi^0,\chi^0;
 b^{\alpha},e^{\beta};  A_{\mu}^{tr},A_{\nu}^{tr}$ have a standard form and for large
$k$ decrease as $k^{-2}$, whereas the mixed propagator
$\varphi^a,\partial_{\mu}A_{\mu}^a$ is a constant $\sim a^{-1}$. The
free field $A_{\mu}$ satisfies the condition
$\partial_{\mu}A_{\mu}^a=0$.

Account of the interaction leads to modification of this condition. The fields
$\varphi^a$ enter the complete lagrangian linearly:
\begin{equation}
\int d^4x\{[\varphi^a [-a \partial_{\mu}A_{\mu}^a+
\frac{g}{2}A_{\mu}^a(\partial_{\mu}\varphi^0-
\partial_{\mu}\chi^0)]- \frac{g}{2}(\varphi^0-\chi^0)A_{\mu}^a \partial_{\mu}\varphi^a\}
 \label{31}
\end{equation}
Variation of this expression with respect to $\varphi^a$ leads to the following
condition on the interacting field $A_{\mu}$
\begin{equation}
(a+ \frac{g}{2}(\varphi^0-\chi^0))\partial_{\mu}A_{\mu}^a=
gA_{\mu}^a(\partial_{\mu}\varphi^0-\partial_{\mu}\chi^0) \label{32}
\end{equation}

Now we shall derive the relations between the Green functions, which follow from the
gauge invariance of the effective action in the eq.(\ref{28}) and replace the usual ST
identities in the present case.

We consider the Green function generating functional given by the integral
\begin{eqnarray}
Z= \int \exp\{i \int[ \tilde{L}(A_{\mu}, \varphi,
\chi,b,e)+\lambda^a(\varphi^a-\chi^a)\nonumber\\
+J_{\mu}^aA_{\mu}^a+\zeta^{\alpha}(\varphi^{\alpha}-\chi^{\alpha})+\xi^{\alpha}(\varphi^{\alpha}+\chi^{\alpha})+\kappa^*b+b^*\kappa+\sigma^*e+e^*
\sigma]dx \}d \mu \label{34}
\end{eqnarray}
where $ \tilde{L}$ is the gauge invariant Lagrangian standing in the exponent of the
integral (\ref{28}), and $J_{\mu}, \zeta, \xi, \kappa, \sigma$ are external sources.
  Let us make the change of variables given by the eq.(\ref{8}). Due to the gauge
  invariance of the Lagrangian $\tilde{L}$ the only terms which change under this
  transformation are the source terms and the gauge fixing term. Using the fact that
  the integral(\ref{34}) does not change under this transformation we get
  \begin{eqnarray}
  \int \exp \{i \int[ \tilde{L}+\lambda^a(\varphi^a-\chi^a)+s.t.]dx \{\lambda^a(y)[
  a+ \frac{g}{2}(\varphi^0(y)-\chi^0(y))]+\nonumber\\ i
  \partial_{\mu}J_{\mu}^a(y)+\zeta^a(y)[a+ \frac{g}{2}(\varphi^0(y)-\chi^0(y))]+ \ldots \}d \mu=0
  \label{35}
  \end{eqnarray}
  Here $s.t.$ stands for the source terms and $\ldots$ denote the variation of all
  remaining source terms. It is convenient to make further redefinitions:
  \begin{equation}
\lambda^a(a+ \frac{g}{2}(\varphi^0-\chi^0))=\lambda'^a
  \label{36}
\end{equation}
\begin{equation}
\varphi^a-\chi^a= (\varphi^a-\chi^a)'(a + \frac{g}{2}(\varphi^0-\chi^0))
 \label{37}
\end{equation}
After such redefinition the eq.(\ref{35}) acquires the form
\begin{eqnarray}
\int \exp \{i \int[\tilde{L}(A_{\mu},(\varphi^a+\chi^a),(\varphi^a-\chi^a)(a+
 \frac{g}{2}(\varphi^0-\chi^0), \varphi^0, \chi^0,
b^{\alpha},e^{\alpha})\nonumber\\+\lambda^a(\varphi^a-\chi^a)+J_{\mu}^aA_{\mu}^a+
\zeta^a(\varphi^a-\chi^a)(a+ \frac{g}{2}(\varphi^0-\chi^0))^{-1}+ \ldots]dx\}\nonumber\\
\{\lambda^a(y)+
\partial_{\mu}J_{\mu}^a(y)+\zeta^a(y)(a+ \frac{g}{2}(\varphi^0(y)-\chi^0(y))+ \ldots\}d\mu=0
\label{38}
\end{eqnarray}
 Here $\ldots$ denote all remaining source terms and their variations under
transformation (\ref{8}).

This equation replaces the standard system of ST identities. In particular the
simplest identities, which follow from the eq.(\ref{38}) are
\begin{equation}
< \lambda^a(x)A_{\mu}^b(y)>=i \partial_{\mu}\delta(x-y)\delta^{ab} \label{40}
\end{equation}
\begin{equation}
<\lambda^a(x)(\varphi+\chi)^b>=0
 \label{41}
\end{equation}
We postpone a detailed analysis of these relations as well as
consideration of the renormalization procedure to a separate
publication.

As follows from our previous discussion the $S$-matrix defined by the eq.(\ref{28})
coincides with the Coulomb gauge scattering matrix given by the eq.(\ref{1}). The
eq.(\ref{1}) strictly speaking defines a unique $S$-matrix for the Yang-Mills theory
only in perturbation theory. Although the eq.(\ref{28}) formally makes sense beyond
the perturbation theory, the proof of unitarity relies on its equality to the Coulomb
gauge $S$-matrix. At present I do not know an independent proof of the unitarity of
the $S$-matrix (\ref{28}).

\section{Discussion.}

The main goal of this paper was to show that Yang-Mils theory allows a manifestly
Lorentz invariant formulation with gauge invariant ghost fields interaction. In this
formulation Yang-Mills theory demonstrates a remarkable similarity to QED. In
particular as in QED the gauge condition (\ref{28}) does not lead to an ambiguity in
the choice of representative in the class of gauge equivalent configurations. The
relations between the Green functions which replace in this case the standard
ST-identities also may be derived in a way similar to QED. Gauge invariance of the
effective action simplifies the construction of invariant regularization and may be
helpful for invariant regularization of non-Abelian supersymmetric models. Finally
this construction may be useful for a nonperturbative analysis of the Green functions
on the basis of Dyson-Schwinger equations having in mind that the gauge condition
(\ref{25}) does not introduce Gribov ambiguity. This problem requires further
investigation.

{\bf Acknowledgements} \\
This work was initiated when the author was visiting Humboldt
University in Berlin. I am grateful to M. Mueller-Preussker for
hospitality and useful discussion and Humboldt Foundation for
support. I also acknowledge a partial support by RBRF, under grant
08-01-00281a, by the grant for support of leading scientific schools
อุ-795.2008.1 and by the RAS program "Theoretical problems of
mathematics".

\end{document}